\documentclass[prb,twocolumn,preprintnumbers,amsmath,amssymb]{revtex4}

\usepackage{amsmath}
\usepackage{graphicx}
\usepackage{dcolumn}
\usepackage{bm}
\usepackage[dvips]{color}

\begin{document}

\title{Negative photoconductance in SiO$_2$(Co)/GaAs heterostructure in the avalanche regime}
\author{L.~V.~Lutsev$^{1}$}
\author{V.~V.~Pavlov$^{1}$}
\author{P.~A.~Usachev$^{1}$}
\author{A.~A.~Astretsov$^{1,2}$}
\author{A.~I.~Stognij$^{3}$}
\author{N.~N.~Novitskii$^{3}$}

\affiliation{$^{1}$Ioffe Physical-Technical Institute, Russian
Academy of Sciences, 194021 St. Petersburg, Russia}
\affiliation{$^{2}$Academic University –-- Nanotechnology Research
and Education Centre, Russian Academy of Sciences, 194021, St.
Petersburg, Russia} \affiliation{$^{3}$Scientific and Practical
Materials Research Centre, National Academy of Sciences of Belarus,
220072, Minsk, Belarus}
\date{\today}

\begin{abstract}
In the avalanche regime we observed the negative photoconductance of
heterostructures of silicon dioxide films containing cobalt
nanoparticles grown on gallium arsenide, SiO$_2$(Co)/GaAs. Light
irradiation with the photon energy less than the bandgap energy of
the GaAs creates holes trapped on defects within the GaAs bandgap,
suppresses the avalanche feedback and causes a reduction of the
current flowing in the SiO$_2$(Co)/GaAs.
\end{abstract}

\maketitle

\email{l_lutsev@mail.ru}

Devices based on avalanche processes such as avalanche
photodetectors, photomultipliers, avalanche transistors are critical
components in high-speed communication systems, optical radars,
quantum cryptography, quantum computing, infrared imaging, laser
ranging \cite{Gisin02,Prev07,Voll10,Nak2000}. Due to the inherent
positive feedback mechanism involved in the impact ionization
avalanche process \cite{Sze}, these devices have high
photosensitivity and the ability to switch very high currents with
less than a nanosecond rise and fall times. Single-photon detection
is one of the most challenging goals of photonics. This goal can be
achieved by use of detectors working in the avalanche regime.
Single-photon avalanche detectors with self-quenching
\cite{Zhao07,Jiang12} show great suppression in excess noise.
Avalanche photodetectors with negative resistance characteristics
exhibits the internal radio-frequency-gain effect -- the enhanced
response in microwave frequencies \cite{Kim03,Kang07}. Another goal
of the avalanche devices such as avalanche transistors and impact
avalanche transit time (IMPATT) devices is generation of
ultra-narrow pulses and high power signals in microwave,
millimeter-wave and terahertz frequencies \cite{Vain10,Ach12}. Thus,
one can conclude that investigation of the avalanche process and
manipulation of the impact ionization is very important for various
applications.

In this Letter we study influence of light irradiation on the
avalanche in (SiO$_2$)$_{40}$Co$_{60}$/GaAs heterostructures and
observe the negative photoconductance in the infrared region -- the
current flowing in the heterostructure decreases under the light
irradiation. One needs to note that the negative photoeffect can be
observed not only on systems with avalanche process, but on
semiconductor structures with quantum wells
\cite{Tut89,Cheng94,Yak2000} and in films with metal nanoparticles
coated by self-assembled monolayers \cite{Nak09}. In the first case,
the effect is due to the electron confinement in well regions. In
the second case, the effect is caused by light-induced creation of
mobile charge carriers whose transport is inhibited by carrier
trapping in transient polaron-like states. Investigation of
SiO$_2$(Co)/GaAs heterostructures is important because the extremely
large magnetoresistance ($10^5$ \%) is observed in these structures
at the avalanche regime at room temperature \cite{Lut05,Lut09}.

{\bf Experiment.} Experiments were performed on metal-dielectric
heterostructures composed of thin film of amorphous silicon dioxide
with cobalt nanoparticles deposited on gallium arsenide substrates
(SiO$_2$)$_{100-x}$Co$_x$/GaAs [the abridge notation is
SiO$_2$(Co)/GaAs]. $n$-GaAs substrates with thickness of 0.4\,mm are
of the (001)-orientation type. Electrical resistivity of GaAs chips
was equal to 0.93$\times 10^5$\,$\Omega\cdot$cm. Prior to the
deposition process, substrates were polished by a low-energy oxygen
ion beam~\cite{Stog02,Stog03}. The roughness height of the polished
surfaces was less than 0.5\,nm. The SiO$_2$(Co) films were prepared
by the ion-beam deposition technique using a composite cobalt-quartz
target onto GaAs substrates heated to 200$^{\circ}$C. The Co
concentration in SiO$_2$ matrix was specified by a relation of
cobalt and quartz surface areas. The film composition was determined
by the nuclear physical methods of element analysis using a deuteron
beam. The cobalt to silicon atomic ratio was measured by the
Rutherford backscattering spectrometry of deuterons. The oxygen
concentration in films was determined by the method of nuclear
reaction with deuterons at $E_d$ = 0.9 MeV: ${ }^{16}$O $+d\to p+{
}^{17}$O. This technique is described in more detail
elsewhere~\cite{Zvon}. For the samples studied, the relative content
of cobalt $x$ is equal to 60\,at.\% and the film thickness is 40 nm.
The average size of Co particles was determined by the small-angle
X-ray scattering and is equal to 3.5 nm. Protective Au layer of a
thickness 3-5 nm have been sputtered on SiO$_2$(Co) films. The Au
layer was used as a contact in experiments.

\begin{figure}
\begin{center}
\includegraphics*[scale=0.39]{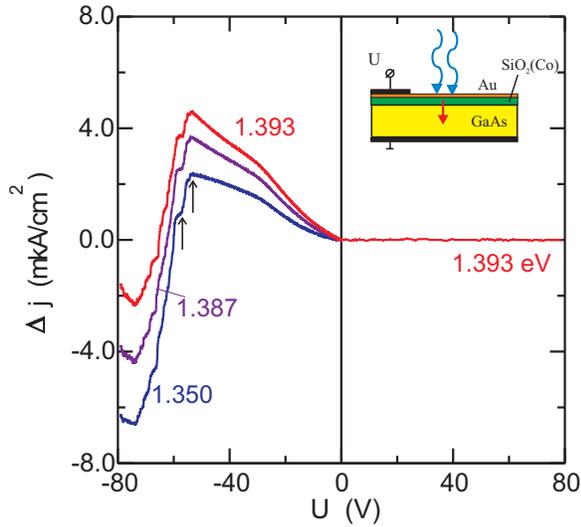}
\end{center}
\caption{ Current change $\Delta j$ caused by the light irradiation
with photon energy $\varepsilon =$ 1.350\,eV, 1.387\,eV and
1.393\,eV versus the voltage $U$ applied on the SiO$_2$(Co)/GaAs
heterostructure. } \label{Fig1}
\end{figure}

In order to measure the the current change $\Delta j$ we use the
lock-in technique with modulation of light beam at the frequency
40~Hz. Fig. \ref{Fig1} shows the current change $\Delta j$ caused by
the linear-polarized light irradiation with photon energy
$\varepsilon =$ 1.350\,eV, 1.387\,eV and 1.393\,eV versus the
voltage $U$ applied on the SiO$_2$(Co)/GaAs heterostructure at room
temperature. One contact was on the GaAs substrate, and the other
(Au contact) -- on the SiO${ }_2$(Co) granular film. The light
intensity $P$ is equal to 2.6~mW/cm$^2$. Photon energies are less
than the GaAs bandgap energy $E$, the depth of penetration of light
into GaAs is high and the light reaches the region of the avalanche
process in the GaAs. The $\Delta j$ dependencies have different
character at positive and negative voltages $U$ applied to the
SiO$_2$(Co) film. At positive voltages no changes in the current is
observed. At small values of negative voltages the current change
$\Delta j$ increases. The avalanche process starts at $|U|=$ 54\,V
and the photocurrent $\Delta j$ begins to decrease.  At higher
voltages ($|U|>$ 60\,Â) the light irradiation leads to the
suppression of the avalanche process and the $\Delta j$ becomes
negative. At voltages $|U|=$ 54\,V and 58\,V the step-like
dependence caused by current filaments is observed~\cite{Ker82}.

\begin{figure}
\begin{center}
\includegraphics*[scale=0.5]{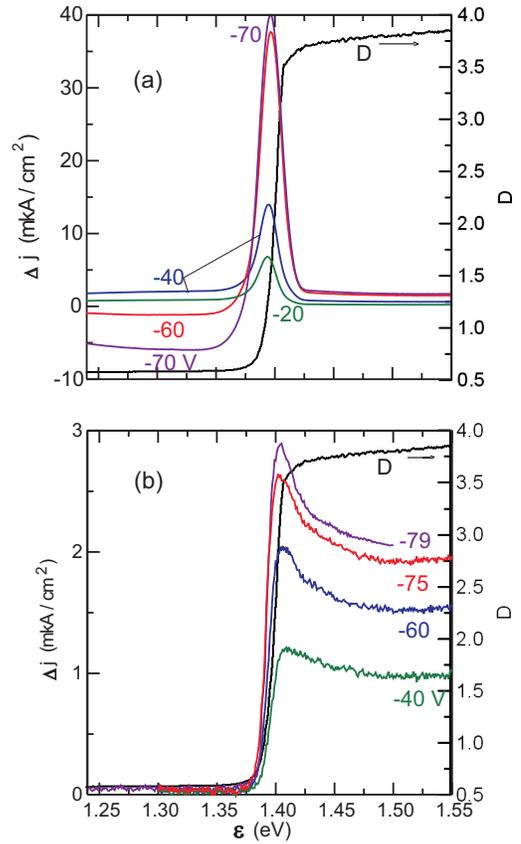}
\end{center}
\caption{(a) Spectral dependence of the current change $\Delta j$
under the light irradiation of the SiO$_2$(Co)/GaAs heterostructure
at negative voltages $U$ on the SiO$_2$(Co) film. (b) Spectral
dependence of the current change $\Delta j$ under the light
irradiation of the GaAs without a SiO$_2$(Co) film. $D$ -- optical
density of samples, $\varepsilon$ -- photon energy. } \label{Fig2}
\end{figure}

Spectral dependencies of the current change $\Delta j$ under the
light irradiation are presented in Fig. \ref{Fig2}(a). At low
negative voltages $U$ applied on the SiO$_2$(Co) film ($U=$ -20,
-40\,V) the $\Delta j$ grows and the positive photoeffect is
observed. The highest growth of the $\Delta j$ exists in the narrow
band of photon energies 1.38 - 1.41\,eV near the bandgap energy $E$
of the GaAs. Outside of this band the growth of the $\Delta j$ is
small. At the avalanche process in the GaAs (negative voltages,
$|U|>$ 54\,V) in the energy band 1.38 - 1.41\,eV the photocurrent
$\Delta j$ retains its growth with voltage increasing. At the same
time, at photon energies $\varepsilon <$~1.38\,eV the decrease of
the current flowing in the heterostructure is observed and at $|U|>$
54\,V the current change $\Delta j<0$. For $\varepsilon >$~1.41\,eV
one can observe sharp decrease of the $\Delta j$, but the
photocurrent retains positive values. Fig.~\ref{Fig2}(b) shows
spectral dependencies of the $\Delta j$ under the light irradiation
of the GaAs without a SiO$_2$(Co) film at negative voltages $U$ on
the GaAs. The light intensity $P$ is equal to 0.3~mW/cm$^2$. In
contrast to SiO$_2$(Co)/GaAs heterostructures, in the GaAs without a
SiO$_2$(Co) film at photon energies $\varepsilon <$~1.38\,eV the
negative photoeffect is absent and at photon energies $\varepsilon
>$~1.41\,eV spectral dependencies of the current change $\Delta j$
do not have the sharp decrease observed on SiO$_2$(Co)/GaAs
heterostructures [Fig. \ref{Fig2}(a)]. We note that maxima of
$\Delta j$ in Fig.~\ref{Fig2}(a) are shifted to lower energies in
comparison to maxima of $\Delta j$ in Fig.~\ref{Fig2}(b).

\begin{figure}
\begin{center}
\includegraphics*[scale=0.37]{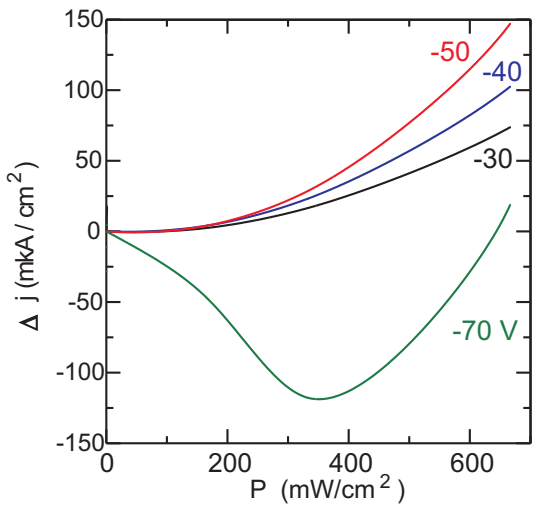}
\end{center}
\caption{Current change $\Delta j$ in the SiO$_2$(Co)/GaAs
heterostructure caused by the light irradiation with photon
wavelength $\lambda =$ 1.050\,$\mu$m ($\varepsilon =$ 1.181\,eV)
versus the light intensity $P$ at different negative voltages
applied on the SiO$_2$(Co) film. } \label{Fig3}
\end{figure}

The photocurrent $\Delta j$ depends on the light intensity $P$
(Fig.~\ref{Fig3}). Without an avalanche in the heterostructure
(negative voltages, $|U|<$ 54\,V), the $\Delta j$ grows with light
intensity increasing. In the avalanche regime ($|U|=$ 70\,V), the
photocurrent changes its sign and the dependence $\Delta j$ versus
$P$ has nonlinear character.

\begin{figure}
\begin{center}
\includegraphics*[scale=0.38]{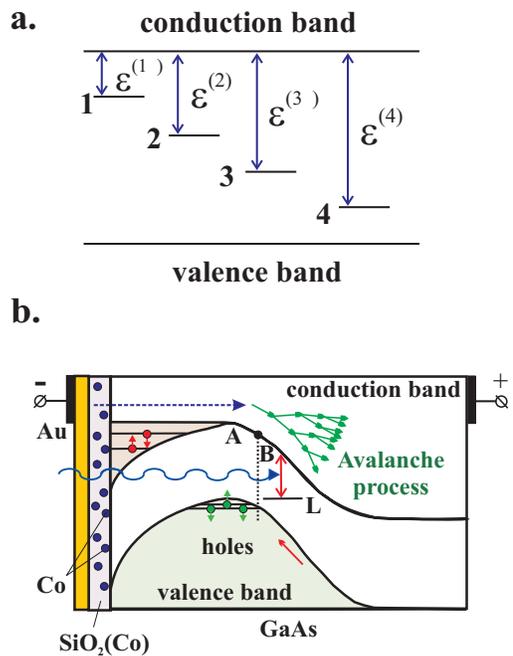}
\end{center}
\caption{ (a) Energy level scheme for the GaAs near the interface.
(b) Schematic band diagram of the GaAs in the SiO$_2$(Co)/GaAs
heterostructure at the applied electrical field in the avalanche
regime. $A$ -- maximum point of the potential barrier, $B$ -- start
point of the impact ionization, $L$ -- energy levels of defects
inside bandgap. } \label{Fig4}
\end{figure}

{\bf Discussion.} Since the interface region of the GaAs contains
oxygen ions leaved after the polished process, then according to
\cite{Lin76,Yu84} in addition to the EL2 defect level there are
oxygen-ion levels in the GaAs bandgap. The four-level model [Fig.
\ref{Fig4}(a)] describes the presence of these levels in the GaAs
\cite{Lin76}. The temperature dependence of dark conductivity near
room temperature is controlled by the thermal excitation of
electrons from level 1 which lies $\varepsilon^{(1)} =$ 0.48\,eV
below the conduction band. The value of $\varepsilon^{(1)}$
corresponds to the activation energy $\varepsilon =$ 0.47\,eV of the
SiO$_2$(Co)/GaAs structure~\cite{Lut05}. In thermal equilibrium
other three levels $\varepsilon^{(2)} =$ 0.74\,eV,
$\varepsilon^{(3)} =$ 1.0\,eV and $\varepsilon^{(4)} =$ 1.25\,eV are
mostly occupied.

The schematic band diagram of the action of light irradiation on the
avalanche process and on the feedback in SiO$_2$(Co)/GaAs
heterostructures is shown in [Fig. \ref{Fig4}(b)]. The
spin-dependent potential barrier is formed in the accumulation
electron layer in the semiconductor near the interface (point
A)~\cite{Lut09,Lut06a}. The impact ionization induced by injected
electrons produces holes, which move and are accumulated in the
region of the potential barrier. Existence of holes in the region of
the barrier lowers the barrier height, grows the electron current
flowing through the barrier and leads to the enhancement of the
avalanche process. Due to the formed positive feedback small
variations in the barrier height give great changes in the current.
Light irradiation of heterostructure leads to a creation of free
electrons in a conduction band and holes in a valence band, as well
as localized electrons and holes on defects $L$ inside bandgap of
GaAs.

To describe photoinduced processes let us consider the current
change caused by light in the avalanche regime. Without a light
irradiation the current $j_0$ flowing in a semiconductor structure,
where the impact ionization process is formed, is the sum of
electron and hole currents, $j_0=j_e+j_h=(1+\beta)j_e$, where
$\beta$ is the positive feedback produced by holes. Influence of the
light irradiation on the current density flowing in a semiconductor
structure can be written as

\begin{equation}
j_t(P,\varepsilon)=
[1+\beta(P,\varepsilon)][j_0+j_c(P,\varepsilon)], \label{eq1}
\end{equation}

\noindent where $j_c$ is the electron current density in a
conduction band induced by the light irradiation, $P$ is the light
intensity, $\varepsilon$ is the photon energy, and $\beta$ is the
function of $P$ and $\varepsilon$. The change in the current density
caused by light is determined by the difference between the current
density $j_t(P,\varepsilon)$ under the light irradiation of the
intensity $P$ and the current density $j_t(0,\varepsilon)$ without
an irradiation

\begin{equation}
\Delta j(P,\varepsilon)= j_t(P,\varepsilon)-j_t(0,\varepsilon).
\label{eq2}
\end{equation}

Due to the positive feedback $\beta$ the current change $\Delta
j(P,\varepsilon)$ can be negative, i.e. negative photoeffect is
observed, and can be positive ($\Delta j(P,\varepsilon)>0$). Taking
into account Eqs. (\ref{eq1}) and (\ref{eq2}), we find that the
negative photoeffect is realized, when the feedback contribution to
the current decrease is greater than the contribution to the current
increase caused by the electron creation in the conduction band and
the inequality

\begin{equation}
\frac{\partial\beta(P,\varepsilon)}{\partial P}<
-\frac{1+\beta(P,\varepsilon)}{j_0+j_c(P,\varepsilon)}\cdot\frac{\partial
j_c(P,\varepsilon)}{\partial P} \label{eq3}
\end{equation}

\noindent fulfills.

The suppression of the avalanche process presented in Fig.
\ref{Fig1} can be explained in the following way. If the photon
energy is less than the bandgap energy $E$, $\varepsilon<E$, light
irradiation causes a creation of conduction electrons in the
conduction band and holes trapped on defects within the bandgap of
GaAs (Fig. \ref{Fig4}). Localized holes on the levels $L$ form the
region of immovable positive charge. This region hinders the
movement of holes, which create in the avalanche process in the
valence band and move to the potential barrier. Consequently, the
positive feedback $\beta(P,\varepsilon)$ decreases. If the value of
$\beta(P,\varepsilon)$ is high to fulfill relation (\ref{eq3}), the
current change $\Delta j$ becomes negative.

For $\varepsilon <$~1.38\,eV (Fig. \ref{Fig2}) the avalanche process
is suppressed, the value of the positive feedback
$\beta(P,\varepsilon)$ decreases and the photocurrent $\Delta j$
becomes negative. For $\varepsilon
>$~1.41\,eV one can observe sharp decrease of the $\Delta j$, but
the photocurrent retains positive values. This decrease is due to
the quantum well formed near the interface [Fig.~\ref{Fig4}(b)].
Electrons which are created by light irradiation of the
SiO$_2$(Co)/GaAs heterostructure are localized in the quantum well.
Since electron-hole pairs are created and accumulate in the
interface region, the light penetration depth in the GaAs has low
values and the light irradiation influence on the positive feedback
$\beta(P,\varepsilon)$ of the avalanche process is insignificant.
This leads to low values of the $\Delta j$.

The developed model of the current change caused by light in the
avalanche regime can explain the dependence of the photocurrent
$\Delta j$ on the light intensity $P$ (Fig.~\ref{Fig3}). Without an
avalanche process ($|U|<$ 54\,V) at photon energies lesser than the
bandgap energy of the GaAs the light irradiation causes a creation
of conduction electrons in the conduction band and holes trapped on
defects within the bandgap. In this case, $\beta(P,\varepsilon)=0$
and the photocurrent is determined by the current
$j_c(P,\varepsilon)$ of activated electrons (Eq. \ref{eq1}). In the
avalanche regime ($|U|=$ 70\,V) at small values of $P$ for the case
of fulfilment inequality (\ref{eq3}) the feedback contribution to
the current decrease is of great values and the current change
$\Delta j$ is negative. At high values of $P$ the feedback
$\beta(P,\varepsilon)$ is suppressed and the $\Delta j$ grows.

In summary, the negative photoconductance is observed in
SiO$_2$(Co)/GaAs heterostructures in the avalanche regime, when the
photon energy is less than the energy bandgap of the GaAs. The light
irradiation creates holes trapped on defects within the GaAs
bandgap. These localized holes hinder the movement of holes created
in the impact ionization process in the valence band and reduce the
avalanche positive feedback. This leads to the observed decrease of
the photocurrent. Thus, in the avalanche regime SiO$_2$(Co)/GaAs
heterostructures demonstrate not only the extremely large
magnetoresistance, but also the negative photoeffect, which can be
used in sensitive infrared detectors.

The authors gratefully acknowledge the assistance of V.M. Lebedev
(PNPI, Gatchina, Leningrad region, Russia) for determination of the
film composition and R.V. Pisarev and A.M. Kalashnikova for useful
discussions. This work was supported by the Russian Foundation for
Basic Research (Project Nos. 10-02-01008, 10-02-00516, 10-02-90023),
the RAS Programs on Spintronics and Nanostructures, the Ministry of
Education and Science of the Russian Federation (project
2011-1.3-513-067-006).

{e-mail: l\_lutsev@mail.ru}


\begin{thebibliography}{77}

\bibitem{Gisin02} N. Gisin, G. Ribordy, W. Tittel, and H. Zbinden, Rev. Mod. Phys. {\bf 74}, 145 (2002).

\bibitem{Prev07} R. Prevedel, Ph. Walther, F. Tiefenbacher, P. B\"ohi, R. Kaltenbaek, Th. Jennewein,
and A. Zeilinger, Nature {\bf 445}, 65 (2007).

\bibitem{Voll10}  M. Vollmer and K.-P. M\"ollmann, {\it Infrared Thermal Imaging: Fundamentals,
Research and Applications}, (Wiley-VCH, Weinheim, 2010).

\bibitem{Nak2000} T. Nakata, T. Takeuchi, I. Watanabe, K. Makita, and T. Torikai, Electron. Lett. {\bf 36}, 2033 (2000).

\bibitem{Sze}  S. M. Sze, {\it Physics of Semiconductor Devices}, 2nd ed. (Wiley, New
York, 1981).

\bibitem{Zhao07} K. Zhao, A. Zhang, Yu-hwa Lo, and W. Farr, Appl. Phys. Lett. {\bf 91}, 081107 (2007).

\bibitem{Jiang12} X. Jiang, M.A. Itzler, K. O'Donnell, M. Entwistle,
and K. Slomkowski, Proc. of SPIE {\bf 8375}, 83750U (2012).

\bibitem{Kim03} G. Kim, I.G. Kim, J.H. Baek, and O.K. Kwon, Appl. Phys. Lett. {\bf 83}, 1249 (2003).

\bibitem{Kang07} H.-S. Kang, M.-J. Lee, and W.-Y. Choi, Appl. Phys. Lett. {\bf 90}, 151118 (2007).

\bibitem{Vain10} S. Vainshtein, V. Yuferev, J. Kostamovaara, and V. Palankovki, Annual Journal of Electronics, 12 (2010).

\bibitem{Ach12} A. Acharyya and J.P. Banerjee, Terahertz Science and Technology {\bf 5}, 97 (2012).

\bibitem{Tut89} G. Tuttle, H. Kroemer, and J.H. English, J. Appl. Phys. {\bf 65}, 5329 (1989).

\bibitem{Cheng94} J.-P. Cheng, I. Lo, and W.C. Mitchell, J. Appl. Phys. {\bf 76}, 667 (1994).

\bibitem{Yak2000} A.I. Yakimov, A.V. Dvurechenskii, A.I. Nikiforov, O.P. Pchelyakov, and A.V. Nenashev, Phys. Rev. {\bf B62}, R16283 (2000).

\bibitem{Nak09} H. Nakanishi, K.J.M. Bishop , B. Kowalczyk, A. Nitzan, E.A. Weiss, K.V. Tretiakov, M.M. Apodaca, R. Klajn, J.F. Stoddart, and B.A.
Grzybowski, Nature {\bf 460}, 371 (2009).

\bibitem{Lut05} L.V. Lutsev, A.I. Stognij, and  N.N. Novitskii, {JETP Lett.} {\bf
81}, 514 (2005).

\bibitem{Lut09} L.\,V. Lutsev, A.\,I. Stognij, and N.\,N. Novitskii, Phys. Rev. {\bf B80}, 184423 (2009).

\bibitem{Stog02} A.I. Stognij, N.N. Novitskii, and O.M. Stukalov, {Tech. Phys. Lett.}
{\bf 28}, 17 (2002).

\bibitem{Stog03} A.I. Stognij, N.N. Novitskii, and O.M. Stukalov, {Tech. Phys. Lett.}
{\bf 29}, 43 (2003).

\bibitem{Zvon} T.K. Zvonareva, V.M. Lebedev, T.A. Polanskaya, L.V. Sharonova, and
V.I. Ivanov-Omskii, {Semiconductors} {\bf 34}, 1094 (2000).

\bibitem{Ker82} B.S. Kerner and V.F. Sinkevich, {JETP Lett.} {\bf
36}, 436 (1982).

\bibitem{Lin76} A.L. Lin, E. Omelianovski, and R.H. Bube, J. Appl. Phys. {\bf 47}, 1852 (1976).

\bibitem{Yu84} P.W. Yu, Appl. Phys. Lett. {\bf 44}, 330 (1984).

\bibitem{Lut06a} L.\,V. Lutsev, J. Phys.: Condensed Matter {\bf 18}, 5881 (2006).

\end{thebibliography}
\end{document}